\documentclass[12pt,preprint]{aastex}

\newcommand{\kms}{km s$^{-1}$}

\slugcomment{Submitted to ApJ}

\shorttitle{The Nuclear Parsecs of NGC~4151}
\shortauthors{Mundell et al.}

\begin{document}


\title{The Nuclear Regions of the Seyfert Galaxy NGC~4151
$-$ Parsec-scale \HI\ Absorption and a Remarkable Radio Jet}


\author{C.G. Mundell\altaffilmark{1}}
\affil{Astrophysics Research Institute, Liverpool John Moores
University, Twelve Quays House, Egerton Wharf, Birkenhead, CH41 1LD,
U.K.}
\email{cgm@astro.livjm.ac.uk}
\author{J.M. Wrobel}
\affil{National Radio Astronomy Observatory, P.O. Box 0, Socorro, NM
87801-0387}
\email{jwrobel@nrao.edu}
\author{A. Pedlar}
\affil{Jodrell Bank Observatory, Macclesfield, Cheshire, SK11 9DL,
U.K.}
\email{ap@jb.man.ac.uk} 
\and
\author{J.F. Gallimore}
\affil{Department of Physics, Bucknell University, Lewisburg, PA
17837; National Radio Astronomy Observatory, 520 Edgemont Road,
Charlottesville, VA 22903} 
\email{jgallimo@bucknell.edu}


\def\fsec{\hbox{$.\!\!^{s}$}}
\def\farcs{\hbox{$.\!\!^{\prime\prime}$}}
\def\HI{\hbox{H\,{\sc i}}}
\def\H2{\hbox{H$_{\rm 2}$}}
\def\cmsq{cm$^{-2}$}
\def\NH{$N_{\mbox{\scriptsize H}}$}
\def\kms{km~s$^{-1}$}
\def\mJyb{mJy beam$^{-1}$}
\def\fdeg{\hbox{$.\!\!^{\circ}$}}
\def\fsec{\hbox{$.\!\!^{s}$}}
\def\arcmin{\hbox{$^\prime$}}
\def\arcdeg{\hbox{$^\circ$}}
\def\lesssim{\mathrel{\hbox{\rlap{\hbox{\lower4pt\hbox{$\sim$}}}\hbox{$<$}}}}
\def\moresim{\mathrel{\hbox{\rlap{\hbox{\lower4pt\hbox{$\sim$}}}\hbox{$>$}}}}

\altaffiltext{1}{Royal Society University Research Fellow}
 

\begin{abstract}

Sensitive high angular and linear resolution radio images of the
240-pc radio jet in NGC~4151, imaged at linear resolutions of 0.3 to
2.6 pc using the VLBA and phased VLA at $\lambda$21 cm, are presented
and reveal for the first time a faint, highly collimated jet (diameter
$\lesssim$1.4 pc) underlying discrete components, seen in lower
resolution MERLIN and VLA images, that appear to be shock-like
features associated with changes in direction as the jet interacts
with small gas clouds within the central $\sim$100 pc of the galaxy.
In addition, $\lambda$21-cm spectral line imaging of the neutral
hydrogen in the nuclear region reveals the spatial location,
distribution and kinematics of the neutral gas detected previously in
a lower resolution MERLIN study.  Neutral hydrogen absorption is
detected against component C4W (E+F) as predicted by Mundell et al,
but the absorption, extending over 3 pc, is spatially and
kinematically complex on sub-parsec scales, suggesting the presence of
small, dense gas clouds with a wide range of velocities and column
densities. The main absorption component matches that detected in the
MERLIN study, close to the systemic velocity (998 \kms) of the galaxy,
and is consistent with absorption through a clumpy neutral gas layer
in the putative obscuring torus, with higher velocity blue- and
red-shifted systems with narrow linewidths also detected across
E+F.  In this region, average column densities are high, lying in the
range 2.7~$\times$~10$^{19}$~$T_{\rm S}$~$<$~$N_{\rm
H}$~$<$~1.7~$\times$~10$^{20}$~$T_{\rm S}$~\cmsq\ K$^{-1}$ ($T_{\rm
S}$ is the spin temperature), with average radial velocities in the
range 920~$<$~V$_{\rm r}$~$<$~1050~\kms.  The spatial location and
distribution of the absorbing gas across component E+F rules out
component E as the location of the AGN (as suggested by Ulvestad et
al.) and, in combination with the well-collimated continuum structures
seen in component D, suggests that component D (possibly subcomponent
D3) is the most likely location for the AGN. We suggest that
components C and E are shocks produced in the jet as the plasma
encounters, and is deviated by, dense clouds with diameters smaller
than  $\sim$1.4 pc.

Comparison of the radio jet structure and the distribution and
kinematics of ionized gas in the narrow line region (NLR) suggests
that shock-excitation by passage of the radio jet is not the dominant
excitation mechanism for the NLR. We therefore favour nuclear
photoionization to explain the structure of the NLR, although it is
interesting to note that a small number of clouds with low velocity
and high velocity dispersion are seen to bound the jet, particularly
at positions of jet direction changes, suggesting that some NLR clouds
are responsible for bending the jet. Alternatively, compression by a
cocoon around the radio jet due to pressure stratification in the jet
bow shock could explain the bright, compressed optical line-emitting
clouds surrounding the cloud-free channel of the radio jet, as
modelled by Steffen et al.

\end{abstract}




\section{Introduction}

Nuclear activity in galaxies is present over a wide range of
luminosities, from the most distant and powerful quasars, to the
weaker Active Galactic Nuclei (AGN) seen in nearby galaxies such as
Seyferts, LINERs and even the Milky Way (e.g., Huchra \& Burg 1992; Ho
Filippenko \& Sargent 1997). The standard model of nuclear activity
involves the release of gravitational potential energy from galactic
material accreted by a supermassive black hole at the galaxy center
and, although radiation from the central engine is detected across the
electromagnetic spectrum, it was the radio emission that first led to
the discovery of powerful AGN (e.g., Baum \& Minkowski 1960; Hazard,
Mackey \& Shimmins 1963; Bridle \& Perley 1984).  Indeed, the presence
of powerful, highly-collimated relativistic radio jets in radio-loud
quasars and radio galaxies, extending well beyond their host galaxy
(e.g., Fanaroff \& Riley 1974; Bridle \& Perley 1984), provided key
evidence for the exhaust material from a black-hole driven central
engine (Scheuer 1974, Blandford and Rees 1974; Bridle et al. 1994;
Urry \& Padovani 1995).

In contrast, radio-quiet quasars and Seyfert galaxies are ten times
more common but 100 to 1000 times weaker at radio wavelengths than
their radio-loud cousins (e.g., Goldschmidt et al. 1999); consequently
nuclear starbursts have been advocated as the primary power source
instead of black-hole accretion (e.g., Fernandes \& Terlevich
1995). However, instrumental improvements and high resolution
radio imaging with the VLA, MERLIN and the VLBA over the last twenty
years have shown increasing evidence for collimated radio emission in
the form of small-scale radio jets that are weak analogues to jets in
radio-loud AGN (e.g. Ulvestad \& Wilson 1989; Wilson 1991; Nagar et
al. 1999; Kukula et al.  1999) and indicate the presence of a central
black hole and accretion disk, at least in some Seyferts. Similarly,
measurements of high brightness temperatures in milliarcsecond
resolution images of Seyfert nuclei with flat radio spectra have also
suggested support for accretion-powered central engines (Mundell et
al. 2000).

Radio jets, although interesting in their own right, also provide a
valuable probe of the interstellar medium close to the central engine,
in particular the putative obscuring torus, which is thought to
surround the central engine, providing fuel for the AGN and
determining the observed optical spectral differences between type 1
(unobscured, viewed pole-on down the torus axis ) and obscured type 2
(viewed edge-on in the torus equatorial plane) (e.g., Antonucci \&
Miller 1985).  Some of the molecular gas in the torus is expected to
be dissociated and ionized by the central UV/X-ray continuum source
resulting in detectable quantities of neutral and ionized gas (e.g.,
Krolik \& Lepp 1989; Pier \& Voit 1995).  Absorption of the radio jet
continuum emission by intervening torus gas can be measured to
determine ionized gas densities via free-free absorption (e.g.,
Ulvestad, Wrobel \& Carilli 1999; Ulvestad et al. 1999), neutral gas
columns via $\lambda$21-cm neutral hydrogen (\HI) absorption (e.g.,
Mundell et al. 1995; Gallimore et al. 1999; Peck \& Taylor 2002) and
molecular gas content via OH absorption (e.g., Hagiwara et
al. 2000). Due to the small size scales of the radio jet and the
obscuring torus, high angular resolution imaging using the VLBA is
required to spatially resolve gas in the absorbing region.

In this context, NGC~4151, with its large quantities of neutral gas
and elongated radio continuum structure, is an ideal candidate for
such a study (see Ulrich 2000 for a comprehensive review of the
properties of the galaxy and its AGN).  NGC~4151 is a Seyfert type 1.5
(e.g., Osterbrock \& Koski 1976) in an almost face-on ($
i$=21\arcdeg), grand-design, weakly-barred spiral host galaxy, Hubble
type (R')SAB(rs)ab (de Vaucouleurs et al. 1991), that contains
significant amounts of \HI\ throughout its two optically-faint spiral
arms, fat weak bar and nuclear region (Davies 1973; Bosma, Ekers \&
Lequeux 1977; Pedlar et al. 1992; Mundell et al. 1995; Mundell \&
Shone 1999; Mundell et al. 1999).  

Radio continuum observations of the strong radio continuum nucleus
show a linear radio structure in the form of a string of knots,
elongated over $\sim$3\farcs5 (230 pc) in average position angle
(P.A.)  $\sim$77\arcdeg\ (Wilson \& Ulvestad 1982; Johnston et
al. 1982; Carral, Turner \& Ho 1990; Pedlar et al. 1993; Mundell et
al. 1995), which is embedded in diffuse emission extending over
$\sim$10\farcs5 ($\sim$680 pc) (Johnston et al. 1982; Pedlar et
al. 1993). Carral et al. (1990) identified five main knots in the
3.5-arcsec jet at 15 GHz, annotating them C1 to C5.  $\lambda$21-cm
MERLIN observations of this central 4\arcsec\ region (Mundell et
al. 1995) revealed localized and marginally-resolved \HI\ absorption,
with a peak column density of $N_{\rm
H}$~$\sim$6$\times$10$^{19}$~$T_{\rm S}$ cm$^{-2}$, against the
component (C4) in the radio jet which is thought to contain the AGN;
no absorption was detected against the other jet components to a
limiting column density of $N_{\rm
H}$~$\sim$2$\times$10$^{19}$~$T_{\rm S}$ cm$^{-2}$.  An east-west
column density gradient was observed and, in combination with UV
column densities and early VLBI images, which showed C4 to consist of
two components (C4E and C4W) separated by $\sim$7~pc, led Mundell et
al. (1995) to suggest that the weaker, western component, C4W,
contains the optical/UV nucleus and thus the
\HI\ absorption is taking place against the first component of the
counterjet (C4E), due to gas in the obscuring torus. Structural and
spectral index information obtained from subsequent radio continuum
VLBA observations of the jet at 1.6 and 5~GHz led Ulvestad et
al. (1998) to suggest a similar model but with the AGN located instead
at the emission peak of C4E (component E after Ulvestad et al. 1998);
requiring an absence of \HI\ absorption against the AGN, this model
predicted \HI\ absorption only against the start of the counterjet,
i.e.  the tail of emission extending eastwards of the peak in C4E
(component F after Ulvestad et al. 1998). 

In this paper we re-open the discussion on the possible location of
the AGN based on new spectral line observations, but whatever the
precise location of AGN, NGC~4151 appears to be unusual in having
neutral gas located close to the nucleus, unlike other Seyferts that
show \HI\ absorption associated with gas located in dust disks on
scales of 100$-$200~pc aligned with the host galaxy disk (e.g., Cole
et al. 1998; Gallimore et al. 1999); in this respect NGC~4151 seems
more similar to Compact Symmetric Objects (CSO) and Steep Spectrum
Core (SSC) objects (e.g., Conway 1996; Peck 1999; Peck \& Taylor
2001).

We present sensitive new, high resolution $\lambda$21-cm continuum
imaging, using the VLBA and phased VLA, of the entire 3\farcs5 radio
jet in NGC~4151 at an angular resolution of 40~mas (2.6 pc), along
with higher resolution images of individual components with angular
resolutions down to 5~mas (0.3 pc).  In addition we present analysis
of the first spatially-resolved images of the associated \HI\
absorbing gas with an angular resolution of 10.5 mas (0.7 pc) and
confirm that the neutral gas, although clumpy, is located across the
whole of component C4W (E+F after Ulvestad et al. 1998).  Section
\ref{observations} details the observational parameters and data
reduction performed; in Section \ref{results} we present the results
of both the sensitive, high resolution continuum imaging of the jet
and the \HI\ distribution and kinematics. In Section \ref{discussion}
we present the arguments for the location of the AGN being associated
with component D and discuss the relationship between the radio jet
and the narrow line region.  We discuss the \HI\ absorption in
comparison with Ly$\alpha$ absorption measurements and present our
conclusions in Section \ref{conclusions}.

As discussed in Mundell et al. (1999) the heliocentric radial velocity
of NGC~4151 is $\sim$1000 \kms, but distance estimates vary depending
on the assumed value of H$_{\rm 0}$ and whether the Virgocentric
correction and the relative velocity of the Local Group with respect
to the Virgo cluster are taken into account.  Distance estimates lie
in the range 10$<$D$<$30.5~Mpc for 50$<$H$_{\rm 0}$$<$100~\kms\
kpc$^{-1}$ and 1000$<$V$<$1523 \kms, but since the uncertainty in the
value of H$_{\rm 0}$ is as large as the Virgocentric correction, for
the purposes of this paper we assume a distance to NGC~4151 of
13.3~Mpc, using H$_{\rm 0}$=75
\kms\ kpc$^{-1}$ and the heliocentric velocity of 998\kms, giving a
linear scale of 0.065~pc~mas$^{-1}$ in the galaxy.

\section{Observations and Data Reduction}
\label{observations}

The observations were obtained with the 10 element Very Long Baseline
Array (VLBA - Napier et al., 1994) plus the 27 antennas of the Very
Large Array (VLA - Thompson et al. 1980) used in phased-array mode at
1.4 GHz, during a 14-hour observing run on 19 March 1996. Phased-array
mode involves the analog sums of the sampled and delayed intermediate
frequency (IF) signals from the VLA antennas along each arm; the
combination of the amplified and filtered signals from the three VLA
arms is then recorded by the VLBI data acqusition system in VLBA
format.  When used in phased-array mode to augment the VLBA, the VLA
offers the equivalent sensitivity of a single 115-m antenna and an
improvement in image sensitivity by a factor of $\sim$2.4 compared
with the VLBA alone.  Although dual circular polarizations (right and
left) were recorded, only the parallel hands (RR and LL) were
correlated. The correlated data, using two bit sampling, consisted of
two intermediate frequencies (IF) each with an 8 MHz bandwidth, 2
polarizations and 512 channels, resulting in a maximum spectral
resolution of 15.6 kHz (3.3
\kms) per channel. The first IF was centered near the redshifted \HI\
line of NGC~4151 at 1415.19 MHz corresponding to the recession
velocity 1100 km s$^{-1}$ (radio convention with respect to the Local
Standard of Rest) while the second IF was shifted by 8 MHz relative to
the center of IF1 (center at 1423.19 MHz) to provide a sensitive
measure of continuum emission. The simultaneous VLA data, obtained
with the array in C-configuration, were observed with a 25 MHz
bandwidth centered at 1419.19 MHz encompassing the total 16-MHz VLBA
bandwidth.

Data editing and calibration followed standard methods\footnote{See
E. W. Greisen \& P. P. Murphy 1998, The AIPS Cookbook, on-line at
http://www.cv.nrao.edu/aips/cook.html.} and used the NRAO\footnote{The
National Radio Astronomy Observatory is a facility of the National
Science Foundation operated under cooperative agreement by Associated
Universities, Inc.} Astronomical Image Processing Software (van
Moorsel, Kemball \& Greisen 1996).  VLBA amplitude scales were
determined from standard VLBA antenna gain tables, maintained by NRAO
staff, and measurements of $T_{sys}$ made throughout the run; the
integrated total flux density of NGC~4151, as measured by the VLA, was
determined in the standard way from observations of 3C286.
Observations of J1310+32 (VLBA antennas) were used to provide a check
for relative amplitude differences between antennas and IFs, while
J1642+39 provided manual pulse and bandpass calibrations. Full phase
referencing was not performed thus allowing the maximum possible
observing time to be spent on NGC~4151, which is bright enough for
self-calibration. However short, regular observations of J1209+41 were
interleaved through NGC~4151 scans with a cycle time of (11+2) minutes
to provide a check on the delay-rate corrections; after imaging and
phase/amplitude self-calibration, the amplitude-only corrections from
J1209+41 were applied to NGC~4151. A single-channel pseudo-continuum
dataset for NGC~4151, which was produced by averaging together the
central 75\% of both IFs, was then imaged and used as the input
starting model for subsequent cycles of self-calibration and data
editing (e.g. Walker 1995).  The $(u,v)$ range covered by the data is
large, 70 k$\lambda$ $-$ 40.8 M$\lambda$ at $\lambda$21 cm, so the
data are sensitive to a wide range of spatial frequencies.  Using a
combination of different data weightings and $(u,v)$ tapers, it is
therefore possible to produce a variety of images at different angular
resolutions in order to show fully the complex structures present in
the jet (Figures
\ref{lowresjet}, \ref{hiresjet}, \ref{jetcompc}).  Continuum images
were formed with angular resolutions from 40~mas (2.6 pc) to 5~mas
(0.33 pc) and are discussed more fully in Section \ref{results}. Due
to the lack of full phase referencing, absolute positions from
J1209+41 were not transferred to NGC~4151 and the R.A. and Dec axes
are therefore provided by the pointing centre assumed correlation
position, derived from earlier MERLIN observations; however, measured
positions for E and D that are very close to those derived from the
fully phase-referenced continuum images of Ulvestad et al. (1998),
with the position of the peak flux density in component E measured,
from our 5-mas uniformly weighted continuum image (Figure
\ref{jetcompc}), to be $\alpha=12^h10^m32^s$\llap{.}5821,
$\delta=39\arcdeg24\arcmin21\farcs060$ compared with
$\alpha=12^h10^m32^s$\llap{.}5822,
$\delta=39\arcdeg24\arcmin21\farcs059$ from Ulvestad et al. (1998).

The additional data editing and corrections derived from the
self-calibration of the continuum emission of NGC~4151 were finally
applied to the NGC~4151 spectral line data and all subsequent
processing concentrated on IF1. The continuum contribution in IF1
(formed from the line-free channels 32-182 and 350-450) was subtracted
in the $(u,v)$ plane using {\sc aips} task {\sc uvlin} and the
corresponding `pure continuum' dataset was formed from the same
line-free channels in the $(u,v)$ plane using {\sc avspc}.
Concentrating on the region of the source where absorption is expected
(i.e., components D, E, F), the pure continuum and line datasets
extracted from IF 1 were then Fourier transformed with uniform and
natural weighting (with 1-mas pixels), deconvolved and added together
to form the corresponding absorption cube
(256~$\times$~256~$\times$~512) with the correct continuum level
adjacent to the line.  It should be noted that although a combination
of data from IF1 and IF2 were used to produce maximum-sensitivity
continuum images of the whole jet, this continuum image is not
suitable for addition to the spectral line dataset (from IF1) to form
optical depth maps because small errors in the bandpass calibration
($<$5\%) between the two IFs produce significant errors in optical
depth spectra (Faison et al.  1998); therefore only continuum emission
from line-free channels in IF1 were used to produce the optical depth
maps.  A full spectral-resolution cube (spectral resolution 3.3 \kms)
with 512 channels was produced, whilst spectral averaging was used to
increase the channel sensitivity in spectral line cubes imaged with
128 and 256 channels. The intermediate spectral-resolution natural
cube with 256 channels (6.6 \kms\ per channel) was used as a
consistency check to confirm the presence of multiple narrow
components detected in the 128- and 512-channel cubes (Figure
\ref{fivespec}).  The more sensitive 128-channel cube (spectral
resolution 13.2 \kms) was imaged and used to construct optical depth
maps (see Figure \ref{opac}), which were created using only absorption
stronger that 1.5 \mJyb\ (3$\sigma$).  In Figure \ref{opac}, structures much smaller
than the beam are an artefact of the data processing and are
compatible with the signal-to-noise of the spectra; trends on scales
larger than or comparable to the beamsize are real. Spectra extracted
from a higher angular resolution, uniformly-weighted cube indicates
clumping of the gas on scales $\sim$6~mas, but the signal-to-noise of
the spectral was insufficient for reliable moment analysis.




\section{Results}
\label{results}

\subsection{$\lambda$21-cm Radio Continuum Emission}

\begin{figure*}
\plotone{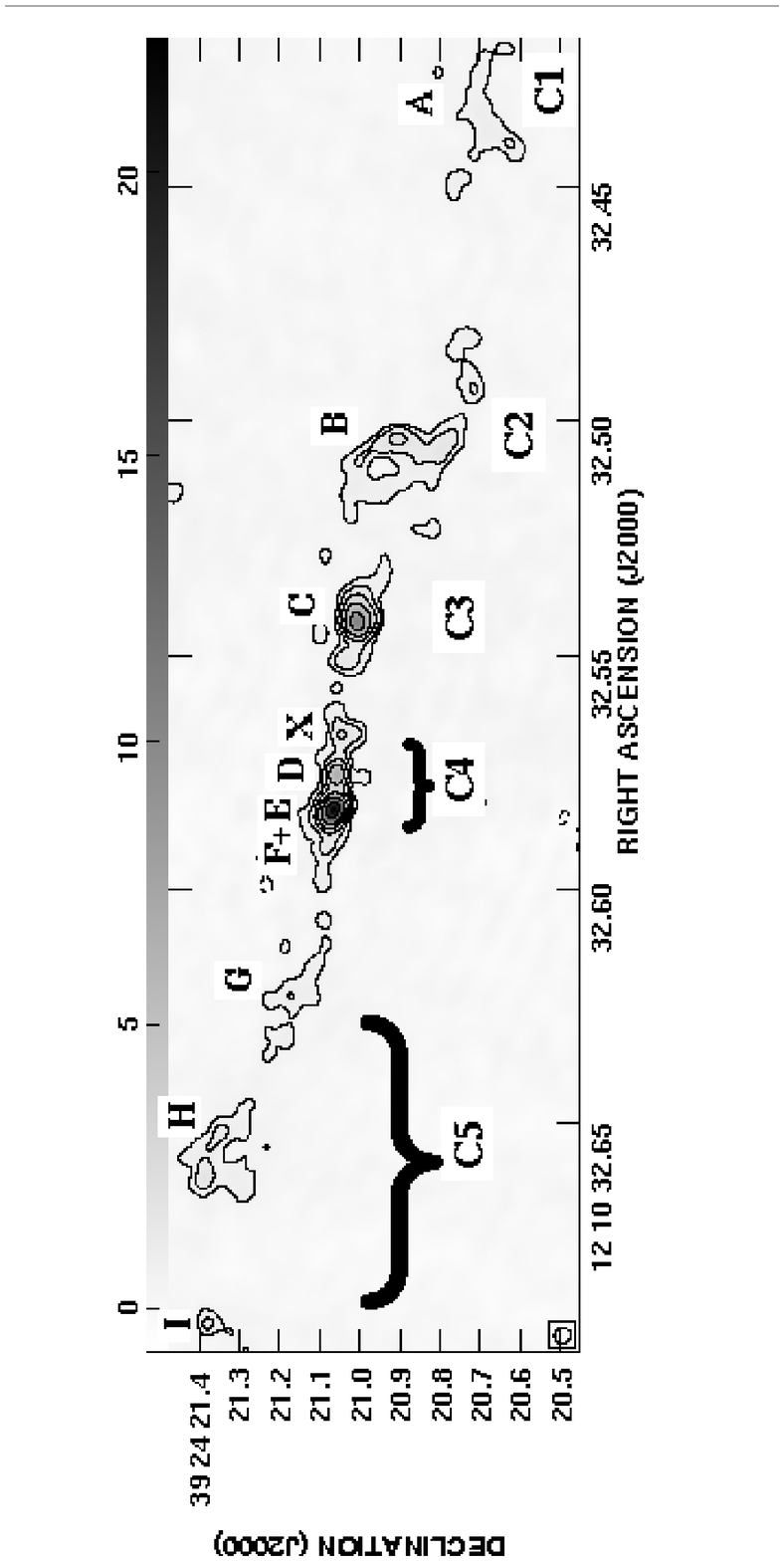}
\caption{Sensitive VLBA+VLA  $\lambda$21-cm
continuum image (produced using 5-M$\lambda$ $(u,v)$ taper), showing
emission along the entire 3\farcs2 (210-pc) length of the radio
jet. Contour levels are (-0.5, 0.5, 1, 2, 4, 8, 16) mJy~beam$^{-1}$
and the 40-mas (2.6-pc) circular restoring beam is shown in the bottom left
corner. Component labeling convention following Ulvestad et al. (1998)
is shown above the jet (A$-$H, plus additional components X, I
identified in this paper) and, for comparison, corresponding component
labeling from lower-resolution studies by Carral et al. (1990) and
Mundell et al. (1995) is shown below the jet (C1$-$C5).
\label{lowresjet}}
\end{figure*}

As discussed in Section \ref{observations}, the $\lambda$21-cm
(1.4-GHz) VLBA+VLA data are sensitive to a wide range of spatial
scales, allowing a variety of images at different angular resolutions
to be produced,  fully probing the structures in the jet.  Figure
\ref{lowresjet} shows continuum emission from the entire 3\farcs2-long
jet ($\sim$210 pc) imaged at 40~mas (2.6 pc) resolution.  For the
first time, weak emission from a highly-collimated radio jet is seen
underlying the five discrete components, or knots (C1$-$C5), seen in
previous lower resolution MERLIN and VLA images (e.g., Carral et
al. 1990; Pedlar et al. 1993). These five knots are now spatially
resolved and can be clearly seen, in the the 25-mas resolution image
(Figure \ref{jetcompc} - upper panel), to coincide with changes in
direction of the jet.

Due to the large number of features visible along the jet in the VLBI
images, the ``C1$-$C5'' labeling scheme used for MERLIN and VLA
images,  as noted by Ulvestad et al. (1998),  is no longer adequate; we
therefore adopt the nomenclature introduced by Ulvestad et al. (1998)
in which the components are labeled from west to east with letters
from A to H.  We identify two new additional components, I (east of H)
and X (between C and D). Table \ref{comps} lists the correspondence
between the component designations used in Mundell et al. (1995) and
Ulvestad et al. (1998).

\begin{figure*}
\plotone{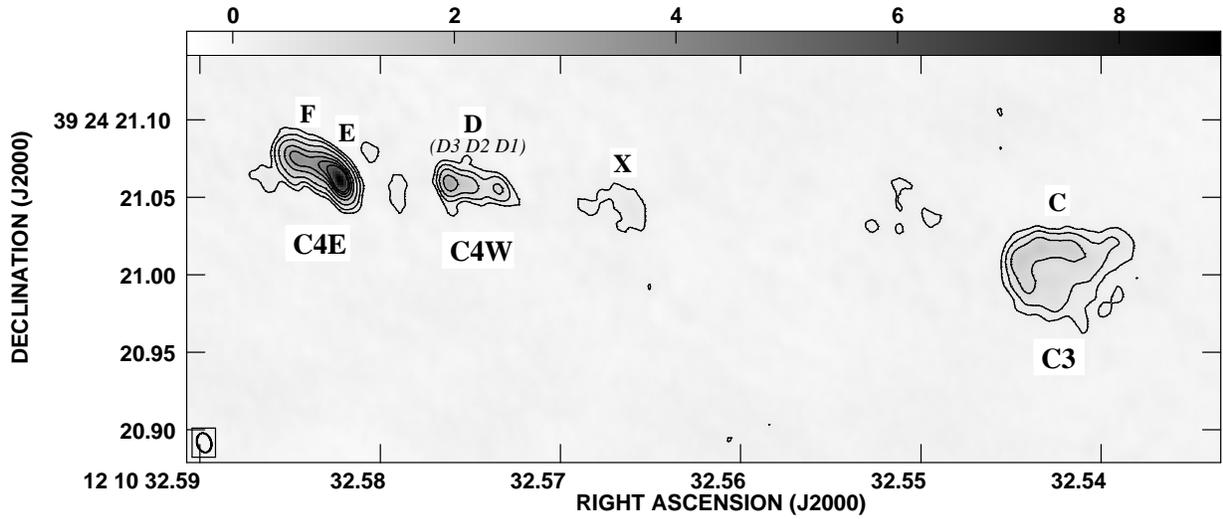}             
\caption{Naturally-weighted VLA+VLBA $\lambda$21-cm continuum image showing
components C, X, D, E, F at high angular resolution. Contour levels
are (-0.25, 0.25, 0.5, 1, 2, 3, 4, 5) mJy~beam$^{-1}$ and the
restoring beam, with size 12.7~$\times$~9.0 mas (0.83~$\times$~0.59 pc)
in P.A. 16\arcdeg, is shown in the lower left corner of the image. As
in Figure \ref{lowresjet}, Ulvestad et al. (1998) component labeling convention
is shown above the jet (plus new component X identified in this paper)
and corresponding component labeling from lower-resolution studies
(e.g., Harrison et al. 1986; Mundell et al. 1995) is shown below the
jet (C3, C4E, C4W).
\label{hiresjet}}
\end{figure*}

\begin{figure*}
\plotone{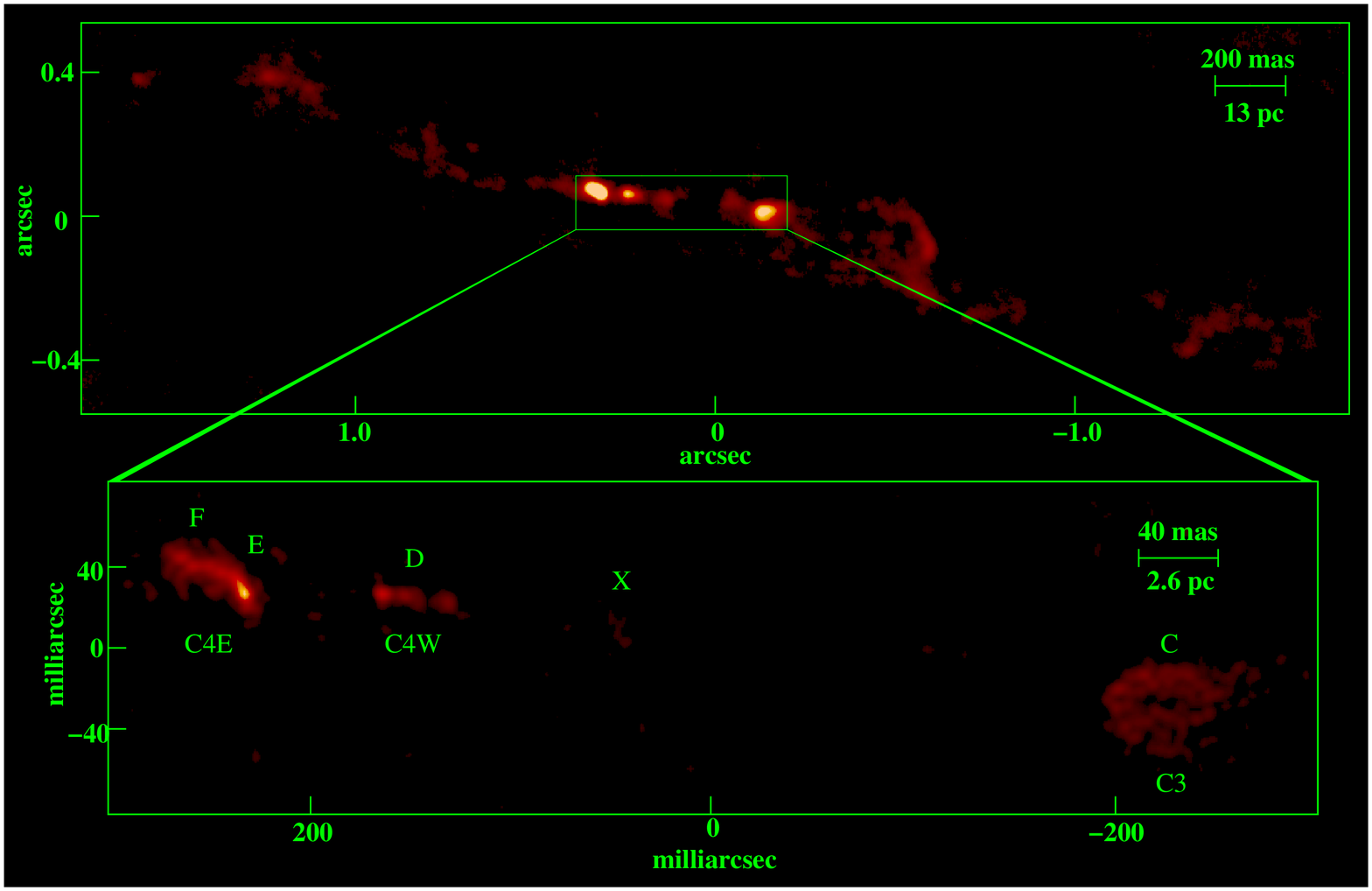}
\caption{Upper panel: full 3\farcs2-length   jet imaged with restoring
beamsize of 25~mas (1.6~pc). Lower panel: maximum angular resolution,
uniformly weighted, image with restoring beamsize of 5~mas (0.33~pc -
1.1 light year), with components marked with Mundell et al. (1995)
convention below and the Ulvestad et al. (1998) convention above.
\label{jetcompc}}
\end{figure*}

Connecting components I and A with a single straight line gives an
average jet P.A. of 77\arcdeg, as found in lower resolution studies
(e.g. Pedlar et al. 1993).  However, the jet structure is more complex
than suggested by a single P.A and, as can be seen in Figures
\ref{lowresjet} and \ref{jetcompc}(a), several sections of the jet
point in a predominantly east-west direction, along
P.A. $\sim$89\arcdeg, i.e. sections connecting components A to the
lower tip of B, F to the lower tip of G and the eastern end of H to I;
finally and most obviously, the peaks of the three subcomponents in D
(Figures \ref{hiresjet}, \ref{jetcompc}(b)) lie along a line in
P.A. $\sim$84\arcdeg\ and connect with X. However, these almost
horizontal sections are not colinear, as the jet exhibits abrupt and
almost step-like, direction changes. These changes are most obvious in
Figure \ref{lowresjet} at components B, G and H, and in Figures
\ref{hiresjet} and \ref{jetcompc}(b) at component E.  The
subarcsecond-scale jet structures therefore do not align with the
average P.A. 77\arcdeg\ and it is clear from Figure \ref{lowresjet}
that the final, `emergent' P.A. for both eastern and western ends of the
jet after it has finished its various deviations is indeed in P.A. $\sim$90\arcdeg.

Although the exact location of the AGN between F and C is not known
(see Section \ref{discussion} for discussion), the jet appears
initially very well-collimated in the vicinity of the components
associated with D (Figure \ref{hiresjet}). Despite deviating quite
abruptly at E, the eastern jet (through components F, G, H, I) remains
more highly collimated along its length than the western jet (through
components C, B, A), which contains more diffuse and extended
components.  No emission is detected in the region between component X
and the eastern tail of C; to a 3-$\sigma$ detection threshold of 0.15
\mJyb, which corresponds to brightness temperature upper limit of
T$_{\rm B}$~$\lesssim$~2.8$\times$10$^5$~K at $\lambda$21 cm in the
40-mas resolution image.

\subsection{Parsec-Scale Neutral Hydrogen Absorption}

Neutral hydrogen absorption is only detected against continuum
components E+F; no \HI\ absorption is detected against component D to
a limiting 3-$\sigma$ absorption depth of 1.5 \mJyb, i.e. a maximum
column density \NH~$\sim$1.1$\times$10$^{19}$~$T_{\rm s}$~\cmsq\ (per
13.2 \kms\ channel), assuming a constant column density across D (see
Section \ref{discussion} for discussion).  Figure \ref{fivespec} shows
some representative absorption spectra taken at five locations
(labelled $\alpha$ to $\epsilon$) across the 3-pc-long continuum
structure of the `banana-shaped' component, E+F; the spectra were
extracted from the 512-channel naturally-weighted cube with angular
resolution 10.5~mas (0.7 pc) and spectral resolution 3.3 \kms.  For
the purposes of moment analysis, we derived the distribution of column
density and corresponding gas kinematics from a spectrally-averaged,
128-channel (spectral resolution 13.2 \kms), naturally-weighted cube
with a restoring beamsize 11.5$\times$9.0~mas (0.75$\times$0.59 pc) in
P.A.  0\arcdeg. Only absorption stronger than the 3-$\sigma$
absorption depth of 1.5 \mJyb\ per channel was included in the moment
analysis.

\subsubsection{Column Densities, Gas Distribution and Kinematics}

\begin{figure*}
\plotone{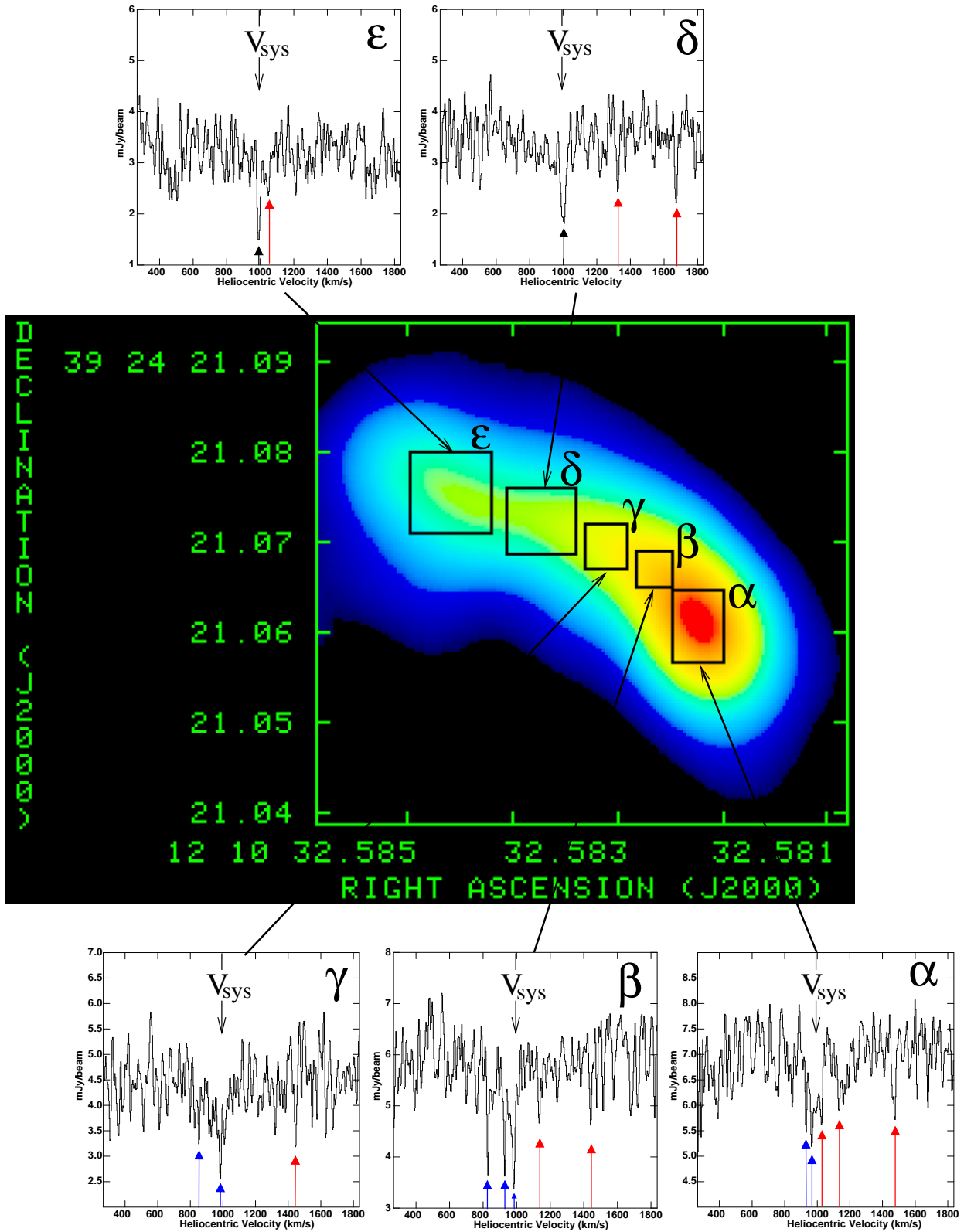}      
\caption{Naturally-weighted 1.4-GHz continuum image with examples of
spectra across this component; the spectra are taken from the
naturally-weighted full spectral resolution (3.3 \kms) data cube and
are displayed with 3-channel (FWHM) Gaussian smoothing
applied. Systemic velocity 998 \kms\ is indicated as V$_{sys}$ and
absorption components are color-coded red (or blue) for positive (or
negative) peak velocities with respect to V$_{sys}$.
\label{fivespec}}
\end{figure*}

\begin{figure*}
\plotone{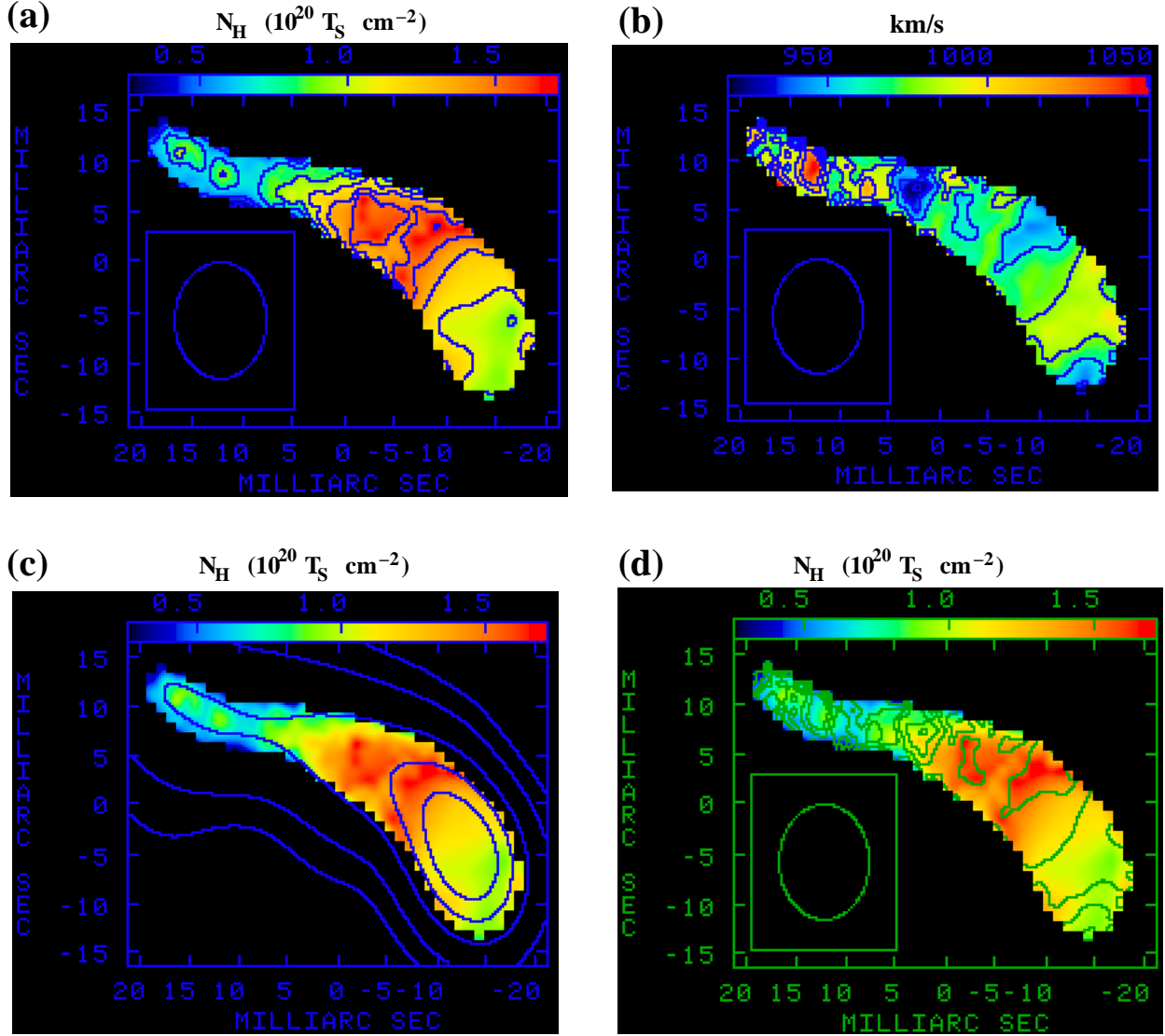}   
\caption{Panels showing moment maps of \HI\ column density (\NH)
and velocity along the `banana' component E+F: (a) \NH\ map with \NH\
contours superimposed; the peak \NH\ is 1.7~$\times$~10$^{20}$~T$_{\rm
S}$~\cmsq (b) velocity map with velocity contours superimposed (c)
\NH\ map with contours of 1.4-GHz radio continuum emission
superimposed with contour levels (-0.8, 0.8, 1.6, 3.2, 4.8, 6.4)~\mJyb\
(d) \NH\ contoured on velocity. The 11.5$\times$9.0-mas
(0.7$\times$0.6 pc) beam (P.A. 0\arcdeg) is shown in the lower left
corner of each image, but omitted from (c) for clarity.
\label{opac}}
\end{figure*}

The distribution of \HI\ column density (\NH) per unit spin
temperature across components E+F is shown in Figure
\ref{opac}(a).  As can be seen in Figures \ref{fivespec} and
\ref{opac}, \HI\ absorption is detected against the whole extent of
E+F (C4W) and is spatially and kinematically complex; across E+F,
average column densities lie in the range
2.7~$\times$~10$^{19}$~$T_{\rm S}$~$<$~$N_{\rm
H}$~$<$~1.7~$\times$~10$^{20}$~$T_{\rm S}$~\cmsq\ K$^{-1}$, where
$T_{\rm S}$ is the spin temperature for which the value is unknown but
is typically $\sim$10$^2$$-$10$^4$ K in the Galaxy (Heiles \&
Kulkarni, 1988) and average radial velocities lie in the range
920~$<$~V$_{\rm r}$~$<$~1050~\kms.  These values are consistent with
the lower resolution (0\farcs16$\times$0\farcs15$\times$26.3 \kms)
MERLIN absorption study in which Mundell et al. (1995) reported an E-W
column density gradient, across the marginally resolved component C4,
3~$\times$~10$^{19}$~$T_{\rm S}$~$<$~$N_{\rm
H}$~$<$~6~$\times$~10$^{19}$~$T_{\rm S}$~\cmsq\ K$^{-1}$ and an
absorption line with Full Width at Half Maximum (FWHM) ~91~\kms\
centered at 998 \kms.  The higher column density values measured in the
VLBA datacube confirms that the absorption is spatially resolved and
occurs on scales significantly smaller than the MERLIN beam.  

The mean column density gradient decreases towards the NE, along the
tail of F, but higher column-density clumps are present throughout
this region, as measured from a higher resolution uniformly-weighted
cube (not shown); in particular, the peak column density \NH\
$\sim$1.7~$\times$10$^{20}$~$T_{\rm S}$~\cmsq\ does not occur at the
location of the continuum emission peak (E) but instead occurs
$\sim$15~mas (1 pc) NE of the southern tip of the `banana' (or
$\sim$0.5~pc from the continuum emission peak). Column densities in
the region of the continuum peak are still significant, but lower at
9~$\times$~10$^{19}$~$T_{\rm S}$~$<$~$N_{\rm
H}$~$<$~1.2$\times$~10$^{20}$~$T_{\rm S}$~\cmsq\ (Figure
\ref{opac}(c)).

Figure \ref{opac}(b) shows the mean velocity field derived from single
Gaussian moment analysis.  Although useful to show the general trend
in velocity across the `banana', this velocity field represents only
the main velocity component close to the galaxy systemic value. As can
be seen in Figure \ref{fivespec}, multiple, narrow velocity components
are present at different positions across the continuum structure. At
this high spectral and spatial resolution one can see a family of
complex absorption lines (e.g. location $\alpha$), multiple velocity
components (e.g. location $\beta$) and narrow single lines
(e.g. locations $\delta$, $\epsilon$) at different locations across
the continuum structure. The velocity components are indicated in
Figure \ref{fivespec} and their properties are listed in Table
\ref{velcomps}.

\section{Discussion}
\label{discussion}

\subsection{The Location of the AGN}

The location of the AGN in NGC~4151 is broadly accepted to be in
component C4 (e.g. Pedlar et al. 1993; Mundell et al. 1995). However,
as discussed previously, it is now known that C4 consists of a number
of subcomponents and the precise association of the AGN with one of
these subcomponents is still a matter of debate. In this study, we use
the characteristics of the radio continuum structure and the \HI\
absorption to argue for the AGN located in component D.

\subsubsection{Radio Continuum Structures}

Components C and (E+F) share some structural similarities having a
`banana'-like ridge on their northern half, but with the tail of (E+F)
pointing eastwards (i.e. counterjet) and that of C pointing westwards
(forward jet); if these opposed arc-like structures are due to
interactions of the radio jet with dense clouds in the circumnuclear
environment, making each component an impact site, their similar
shapes and opposing directions suggest that the AGN lies somewhere
between the two.  The greater flux density of E+F suggests that the AGN
is located closer E+F than to C, or that C is an older component.  In
addition, as described in Section \ref{results}, the continuum
structure of component D shows a high degree of collimation; indeed
the structure resembles radio core-jet structures seen in higher power
AGN such as radio loud quasars.  If the AGN corresponds to one of the
components in D (possibly D3), as originally suggested by Mundell et
al. (1995), this would explain the very linear, core-jet-like,
structure of D and the greater brightness of E+F relative to C.

Assuming that D delineates the well-collimated jet before it has
suffered any interference, a jet diameter of approximately 1.4 pc is
inferred. A small dense gas cloud that wanders into the path of the
radio plasma flow would disrupt or deviate the jet flow. The amount of
disruption depends on the impact parameter of the jet with the cloud,
assuming the cloud diameter is less than that of the jet ($<$1.4
pc). Component E+F can then be understood as a grazing encounter with
a small dense cloud, so the (counter-)jet is deflected at E but
continues to flow over the cloud to F and outwards through G, and
retaining some degree of collimation through to I.

To the west of D, the similar but more diffuse, but clearly shell-like
structure of component C (Figure \ref{hiresjet}) can be explained by
the jet flow to the south-west of the AGN impacting a cloud more
directly and so being more disrupted. With a cloud diameter smaller
than the jet diameter, the jet flow is not terminated at C but
continues to flow around the cloud and onwards to components B and A,
but with significantly more disruption and poorer collimation than its
counterpart to the north-east. Indeed, the flattening of the northern
half of the shell structure of C might be explained by the containment
effect of a dense bounding cloud to the north of C. As shown by HST
studies of ionized gas clouds in the NLR of NGC~4151 (e.g., Wing\'e et
al. 1997; Hutchings et al. 1998; Kaiser et al. 2000), a large number
of gas clouds exist in the inner regions of the galaxy, so the chance
of the radio jet intercepting and being affected by gas clouds is
high.  In addition, the relatively flat radio spectrum of component E
might be explained by re-acceleration of synchrotron-emitting
electrons in the shock produced by the interaction with the cloud,
rather than the AGN-interpretation suggested by Ulvestad et
al. (1998).

\subsubsection{Distribution of \HI\ Absorption }

Mundell et al. (1995) originally suggested that the discrepancy
between the high column densities and systemic gas velocity inferred
from MERLIN radio \HI\ absorption measurements (peak \NH\
$\sim$6.0~$\times$10$^{21}$~\cmsq\ at $\sim$998 \kms\ for ~$T_{\rm
S}$=100 K ) and much lower columns inferred from UV observations of
Ly$\alpha$ absorption (Kriss et al. 1992) with significant blueshifts
(6~$\times$~10$^{17}$~$<$~\NH\ $<$~6~$\times$~10$^{20}$~\cmsq) could
be interpreted as support for the presence of the putative obscuring
torus invoked in Unification Schemes (e.g. Antonucci \& Miller
1985). In this interpretation, the lower column densities of
blueshifted gas seen in the UV are interpreted as outflowing ionised
gas seen along the line of sight to the AGN; the higher column
densities and lower gas velocity seen in the radio are interpreted as
absorption through gas in the obscuring torus along the line of sight
towards the first component of the counterjet. Component C4W (or D)
was then assigned as the true optical/UV nucleus with C4E (E+F)
representing the first radio component in the counterjet against which
the \HI\ absorption was occurring (Mundell et al. 1995).

Ulvestad et al. (1998) used the VLBA to image, at high angular
resolution, the bright components (C, D, E, F) at $\lambda$18 cm and
$\lambda$6 cm and set upper limits for the emission at $\lambda$2
cm. They revised the model of Mundell et al. by suggesting an
alternative location for the AGN in E (as discussed above) and
predicted that \HI\ absorption would only be detected against
subcomponents of F (F3$-$F5 in Ulvestad et al. 1998), with
subcomponents E$-$F2 lying behind fully ionized gas and therefore
resulting in no \HI\ absorption against these components. This
prediction is easily testable with the present data and found to be
inconstent; \HI\ absorption is detected across the whole extent of E+F
(see Section \ref{results} and Figures
\ref{jetcompc}, \ref{fivespec}), ruling E out as the location for the
AGN.  

\begin{figure*}
\plotone{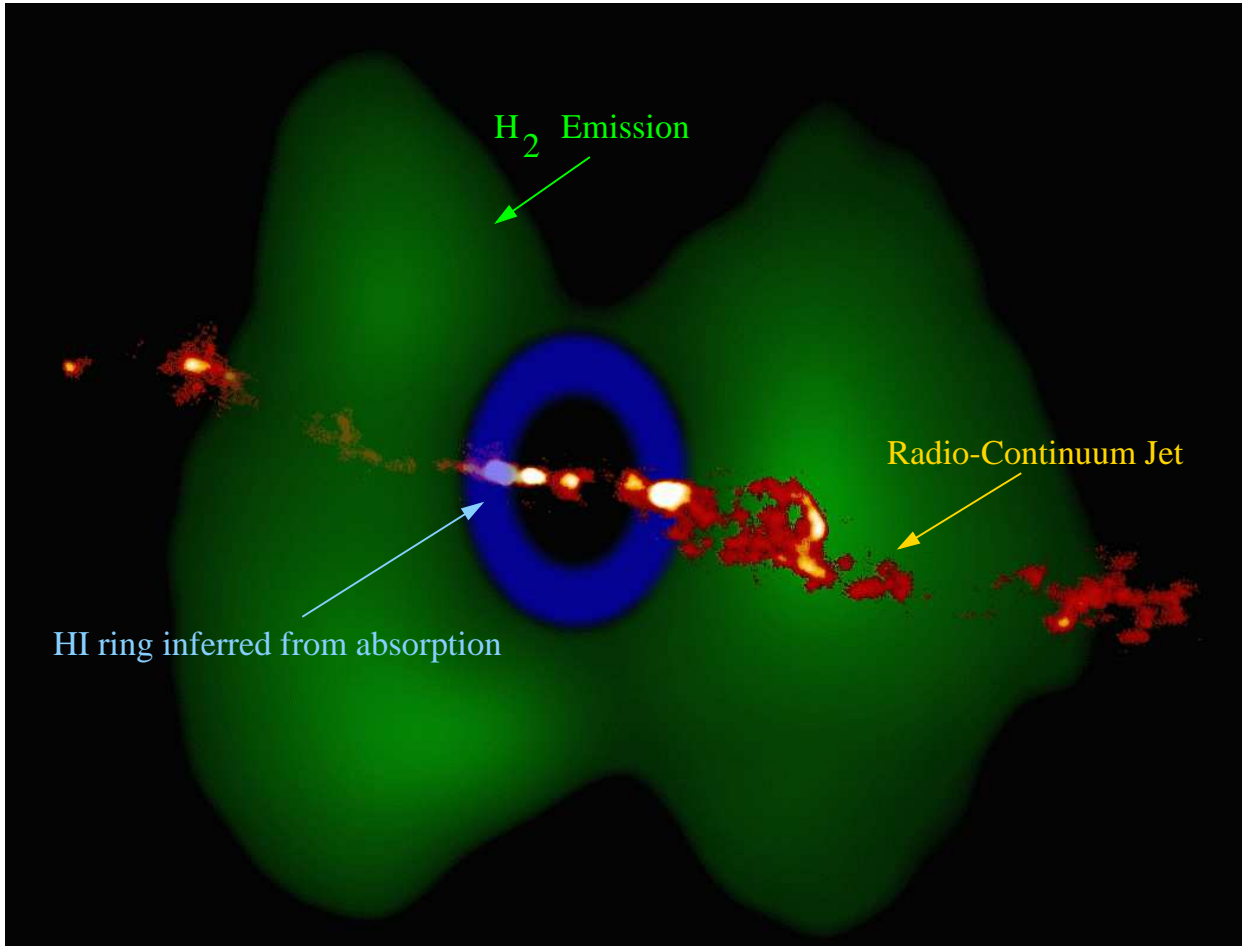}
\caption{Montage of the inner 250-pc of NGC~4151, showing torus of
\H2\ emission in green (from Fernandez et al. 1998),  ring of \HI\ inferred
from absorption measurements in blue and 1.4-GHz radio continuum
emission from radio jet in red. Ionized gas (black) is assumed to fill
the torus inside the \HI\ ring. The segments missing to the north and
south of the \H2\ torus are due to the limited filter width that
excluded the high velocity wings of the \H2\ line; this provides
evidence for rotation of the torus as the northern and southern
segments contain the gas with the highest radial velocities (i.e. line
wings) if the torus is rotating.
\label{montage}}
\end{figure*}

Considering both the \HI\ absorption results and the radio continuum
structures (as discussed above), component D remains the most likely
location for the AGN.  Assuming an ionized/neutral/molecular
`onion-skin' model for the torus, component D would then lie behind
the ionized inner skin of the torus with E+F lying behind the neutral
(dissociated) gas layer and molecular gas would be further out. The
geometry of such a model is illustrated in Figure \ref{montage}. In
this model, the weak diffuse component X, which is resolved out at
higher resolutions and radio frequencies, has been placed at the
centre of the torus since it also lies along the collimation axis
defined by D and so could be an alternative location for the AGN; the
jet can be shifted relative to the H$_{\rm 2}$ image (relative
astronometry is poor) to place whichever radio component contains the
AGN at the torus centre, with the condition that component E+F must
lie behind the neutral gas layer. If subcomponent D3 contains the AGN
and ionized gas fills the inner region of the torus between D3 and the
neutral gas layer starting at E, the radius (R$_{\rm I}$) of the
ionized gas sphere interior to the neutral layer would be
approximately R$_{\rm I}$~$\sim$3.3(sin$^{-1}$$\theta$)~pc (where
$\theta$ is the angle between the jet axis and the line of sight). For
an ionizing flux of 1.5$\times$10$^{54}$ photons per second (Kriss et
al. 1995; Ulvestad et al. 1998), the electron density in Str\"omgren
radius R$_{\rm I}$~$\sim$3.3(sin$^{-1}$$\theta$)~pc is n$_{\rm
e}$$\sim$3.8$\times$10$^{4}$~(sin$^{3/2}$$\theta$)~$f^{-0.5}$~cm$^{-3}$
(where $f$ is the filling factor); this electron density is
intermediate between the densities expected for Seyfert narrow-line
regions ($\sim$10$^3$~cm$^{-3}$) and broad-line regions
($\sim$10$^9$~cm$^{-3}$) and similar to values inferred from free-free
absorption in Mrk~231 and Mrk~348 (Ulvestad et al. 1999) and thermal
bremsstrahlung emission upper limits in NGC~4388 (Mundell et
al. 2000).

Sensitive radio imaging over a range of wavelengths to derive radio
spectra of the subcomponents in D would help to identify a
flat-spectrum, high brightness temperature component indicative of an
AGN and characterise the quantity of ionized, free-free absorbing
foreground gas.  Alternatively, the nucleus itself may not be visible
at centimetric radio wavelengths if significant quantities of ionized
gas lie along our line of sight and free-free absorb radio photons
(e.g. the absence of $\lambda$21-cm continuum emission between
components X and C can be seen in Figure \ref{lowresjet} and is
described in Section \ref{results}); sensitive, high frequency radio
observations, radio proper motion studies and improved HST and VLBA
relative astrometry will be important in determining the true location
of the AGN.

\subsection{The Relationship between the Narrow Line Region and the Radio Jet}

Ionized gas in the narrow line region of NGC~4151 is distributed in a
clumpy, biconical structure, extending over $\sim$2\arcsec\ north-east
and south-west of the bright nucleus, and has been modelled as gas
outflowing at a constant velocity of $\sim$350 \kms\ along the surface
of a bicone with total opening angle of 80\arcdeg, viewed at
10\arcdeg\ outside the cones (Hutchings et al. 1998). The blueshifted
gas southwest of the nucleus is approaching (see also Schultz 1990;
Evans et al. 1993; Robinson et al. 1994; Boksenberg et al. 1995).

A key question in the study of the energetics and kinematics of gas in
NLR regions of Seyferts is the importance of the radio jet.  Competing
models to explain optical line emission from the NLR advocate two main
excitation mechanisms; direct photoionisation by the central continuum
source (e.g., Binette et al., 1996) or shock ionisation caused by the
passage of the radio jet through the ISM (e.g., Dopita \& Sutherland,
1995, 1996; Bicknell et al., 1998). Wedge-shaped morphologies observed
in some NLR's (e.g., Evans et al., 1991; Wilson et al., 1993; Wilson
\& Tsvetanov, 1994; Simpson et al., 1997) are consistent with
photoionisation by a cone of UV radiation from the central
engine. Alternatively, the correlation between radio power and [O{\sc
iii}] luminosity and line-width (de Bruyn \& Wilson 1978; Wilson \&
Willis 1980; Whittle, 1985, 1992), and the approximate spatial
coincidence between radio and optical structures in other NLRs suggest
that the radio jets may be responsible for some of the optical line
emission.

Radio observations of powerful quasars and radio galaxies have shown
evidence for bulk relativistic (non-thermal) flows in the radio jets
(e.g. Pohl et al. 1995; Hough, Zensus \& Porcas 1996 ), but it remains
controversial whether ejected radio plasma in Seyfert galaxies has
bulk relativistic motion (e.g. Ulvestad et al. 1999; Brunthaler et
al. 2000).  Bicknell et al. (1997) suggest that Seyfert jets may be
mildly relativistic close to the black hole and become increasingly
dominated by thermal plasma on larger scales, as they entrain material
from the ISM. They argue that the energetic importance of Seyfert
radio jets has been underestimated and that jets may provide the
energy input to drive the optical emission in the NLR, with jet energy
fluxes (based on a study of NGC~1068) being greater than previously
supposed. Nevertheless, a direct causal relationship between the radio
jet and NLR emission has not yet been established and the relative
contributions of direct photoionisation of emission line regions by
the central continuum source, and strong interaction of the
emission-line gas with radio plasma remains controversial.

Until now, the relatively small angular extent ($\sim$ few arcsecs) of
Seyfert radio jets has limited the detail visible in VLA and MERLIN
images and, radio images at even $\sim$0\farcs1 angular resolution,
have proven insufficient to identify a clear association between NLR
clouds and radio jet components (e.g. Mundell et al. 1995; Kaiser et
al 2000). Now however, the improved sensitivity of new radio VLBI
studies is revolutionising the study of Seyfert radio jets. Figure
\ref{FOCjet} shows our 40-mas resolution image  of the radio jet in
NGC~4151 (from Figure \ref{lowresjet}) overlaid on the F501N
narrow-band image taken with the Faint Object Camera (FOC) on board
the Hubble Space Telescope (HST) (image courtesy of R. Catchpole; see
also Wing\'e et al. 1997). The FOC image with 0\farcs0143 pixels has
been resampled to match the 0\farcs0175 pixel size of this radio image
and, since relative astrometry between the HST and VLBA images is
insufficient for accurate `blind' alignment, the peak of the optical
nuclear emission (assumed originate from the AGN) was aligned with the
peak of radio emission in component D, assumed to contain the AGN. A
reliable comparison of the projected location of the optical
line-emitting clouds and radio jet features is now possible, although
any physical association along the line of sight is still unknown.

The misalignment of the radio jet and NLR supports photoionization of
the NLR by the AGN as the dominant excitation mechanism. A similar
misalignment is present between soft X-ray emission imaged with
Chandra (Yang, Wilson \& Ferruit 2001) and the radio jet. However, a
small number of bright [O{\sc iii}]-emitting clouds appear to be
closely associated with the jet, lying close to and appearing to bound
the radio jet emission; these clouds are labelled 6, 9, 11, 12, 15,
17, 19, 20 in Figure
\ref{FOCjet}. Kaiser et al. (2000) classify all of these clouds as low
velocity ($|v|$~$<$~400~\kms) but high velocity dispersion ($\Delta v
> $~130~\kms) and suggest that they may be intrinsically high-velocity
clouds with smaller projected components of velocity along the line of
sight. In addition, Kaiser et al. (2000) note that clouds 15, 19 and
20 also have high velocity, blueshifted clouds ($|v|$~$>$~1000~\kms)
associated with them that are too faint to be visible in their
labelled WFPC2 images. It is therefore likely that the radio jet might
have cleared a channel through the NLR and suffered disturbance as a
consequence; remaining NLR clouds then act to gently deviate the
plasma flow. Alternatively, as modelled by Steffen et al. (1997), the
enhanced optical emission from the clouds bounding the radio jet might
arise due to compression by a cocoon around the radio jet; a structure
of cylindrical nested shells are produced around the radio jet due to
pressure stratification in the jet bow shock and the brightest
compressed clouds then surround the cloud-free channel of the radio
jet (see Figure 2 in Steffen et al. 1997). Numerical simulations, by
Steffen et al., of this jet propagating through an inhomogeneous
NLR-like medium show optical linewidths as high as 500~\kms, and large
variations in the cloud radial velocities, qualitatively similar to
those observed for NGC~4151.

\subsection{Complex \HI\ Absorption and the Obscuring Torus}

\subsubsection{Comparison with UV/X-ray Absorption}

Absorption by gas along the line of sight to AGN is observed at a
variety of wavelengths, with X-ray and UV absorption occurring
predominantly in ionized gas and radio absorption in neutral and
molecular gas.  The nature of the ``warm'' (or ionized) absorbing gas
detected in X-ray spectra of the nuclei of nearby AGN is still
unknown. As many as 50\% of nearby AGN show absorption by ionized gas
(e.g. George et al. 1998) with many also show associated UV absorption
by highly ionized species such as C {\sc iv} (Crenshaw et al. 1999;
Kraemer et al. 2001). In NGC~4151 broad, saturated Ly$\alpha$
absorption lines are detected in the UV (Kriss et al. 1995; Espey et
al. 1998) and, when considered along with the properties of strong
absorption lines due to other species such as N{\sc v} C{\sc iv} Si
{\sc iv}, indicate that the absorber does not fully cover the source
of background emission (Kraemer et al.  2001). The ionized gas appears
to be outflowing from the nucleus with radial velocities as high as
$-$1680 \kms\ (Kraemer et al. 2001) and cannot account for the
observed X-ray absorption (e.g. George et al. 1998; Yang et
al. 2001). Kraemer et al. (2001) suggest a model in which the UV
absorber covers the BLR, with the X-ray absorber closer to the central
source and obscuring a smaller X-ray emitting region.

The \HI\ column densities in NGC~4151 derived from Ly$\alpha$
absorption (e.g. Kriss et al. 1992, 1995; Espey et al. 1998),
$\sim$3~$\times$10$^{15}$$<$\NH(Ly$\alpha$)~$<$6$\times$10$^{20}$~\cmsq,
are significantly lower than those derived from the present
$\lambda$21-cm observations (see Section \ref{results}), and although
the UV continuum is known to be highly variable, the UV-measured
column densities remain lower than the radio values.  This confirms
that the neutral gas obscuring the radio continuum emission is
distinct from the UV-absorbing gas and we re-propose our geometrical
model (Figure \ref{montage}) in which the $\lambda$21-cm \HI\
absorbing gas is located in a neutral, photo-dissociated layer of the
obscuring torus, detected in front of the first radio component in the
radio counterjet (E+F), while the UV-absorbing gas is outflowing,
ionized gas (possibly blown from the torus walls) along the line of
sight to the nucleus.

\subsubsection{A Clumpy Torus?}

The total absorption spectrum across E+F is consistent with that
detected in the MERLIN study (Mundell et al. 1995), but the
presence of additional narrow absorption lines with higher velocities
(Figure \ref{fivespec}, Table \ref{velcomps}) across E+F argues for
dense clumps of gas within the circumnuclear disk or torus, and may
indicate turbulence within the gas.  Clumpiness of torus gas is
predicted theoretically (e.g. Krolik \& Begelman 1988), with cloud
covering factors close to or greater than unity, and recent X-ray
determinations of column density variations of 20\%$-$80\% over
timescales of months to years (e.g. Risalti, Elvis \& Nicastro 2002)
in Seyfert 2 nuclei support clumping of torus material on scales less
than 1 pc.  Clouds drifting across the background continuum source
would then explain observed column density variations. Sub-parsec
scale radio monitoring of the \HI\ absorption across component E in
NGC~4151 is required to determine if the neutral gas columns vary,
and might therefore be associated with drifting cloud cores, or remain
constant, suggesting \HI\ it is the primary torus volume filler in
which denser clouds are embedded.

\subsubsection{NGC~4151 as a Nearby Radio Galaxy?} 

A similar scenario to that suggested for NGC~4151 has been proposed
for \HI\ absorption detected in the nuclear regions of the compact
symmetric object 1946+708 (Peck, Taylor \& Conway 1999), in which
complex \HI\ absorption is detected against the counterjet (see also
Taylor 1996; Taylor et al 1999).  Indeed the higher rate of detection
of \HI\ absorption in AGN with two-sided, non-boosted radio jets such
as compact symmetric objects (CSO) and compact steep spectrum objects
(CSS) compared with core-dominated radio sources, with Doppler-dimmed
counterjets, provides support for the Unification Scheme in which
objects seem close to the plane of the sky have a more favourable
geometry and orientation for the detection of HI absorption against
the counterjet due to gas in a nuclear obscuring torus (e.g. van
Gorkom et al. 1989; Pihlstr\"om 2001). NGC~4151 is an unusual Seyfert
in having neutral gas located close to the nucleus, unlike other
Seyferts that show \HI\ absorption associated with gas located in dust
disks on scales of 100$-$200~pc aligned with the host galaxy disk
(e.g., Cole et al. 1998; Gallimore et al. 1999); in this respect, the
\HI\ absorption, and indeed the well-collimated radio jet in NGC~4151,
closely resemble parsec-scale structures in CSO and CSS.

\section{Conclusions}
\label{conclusions}

We have used the VLBA and phased VLA at $\lambda$21 cm to study the
parsec-scale radio continuum emission and distribution and kinematics
of neutral hydrogen in absorption in the Seyfert galaxy NGC~4151.  We
find:
 
\begin{itemize}
\item A faint, highly collimated radio jet, diameter $\lesssim$1.4 pc,
underlying brighter radio knots, seen in lower resolution MERLIN and
VLA images, that appear to be shock-like features associated with
changes in direction as the jet interacts with small gas clouds within
the central $\sim$100 pc of the galaxy.
\item H {\sc i} absorption across the whole of component C4W (E+F), as predicted
by Mundell et al. (1995),  with complex spatial and kinematic structure.
\item The location and distribution of the HI absorption and the
structure of the radio continuum emission argue for the AGN being
located in component D, rather than component E as argued by Ulvestad
et al. (1998).
\item We suggest the absorbing gas lies in a thin
photo-dissociated layer of clumpy neutral gas in between the molecular
and ionized gas in the circumnuclear torus, with an inner radius
$\sim$3.3 (sin$^{-1}$$\theta$) pc.
\item The location of the absorbing gas close to the AGN is unusual
for a Seyfert galaxy and we argue that the \HI\ and continuum
properties of NGC~4151 are similar to those seen in compact symmetric
objects and compact steep spectrum objects.
\item Spatial association of the radio jet knots and optical
line-emitting clouds with high velocity dispersion in the narrow line
region suggest some interaction between the radio jet and the clumpy
interstellar medium, but photoionization remains the dominant
excitation mechanism for the NLR.

\end{itemize}

\acknowledgments

\begin{figure*}
\plotone{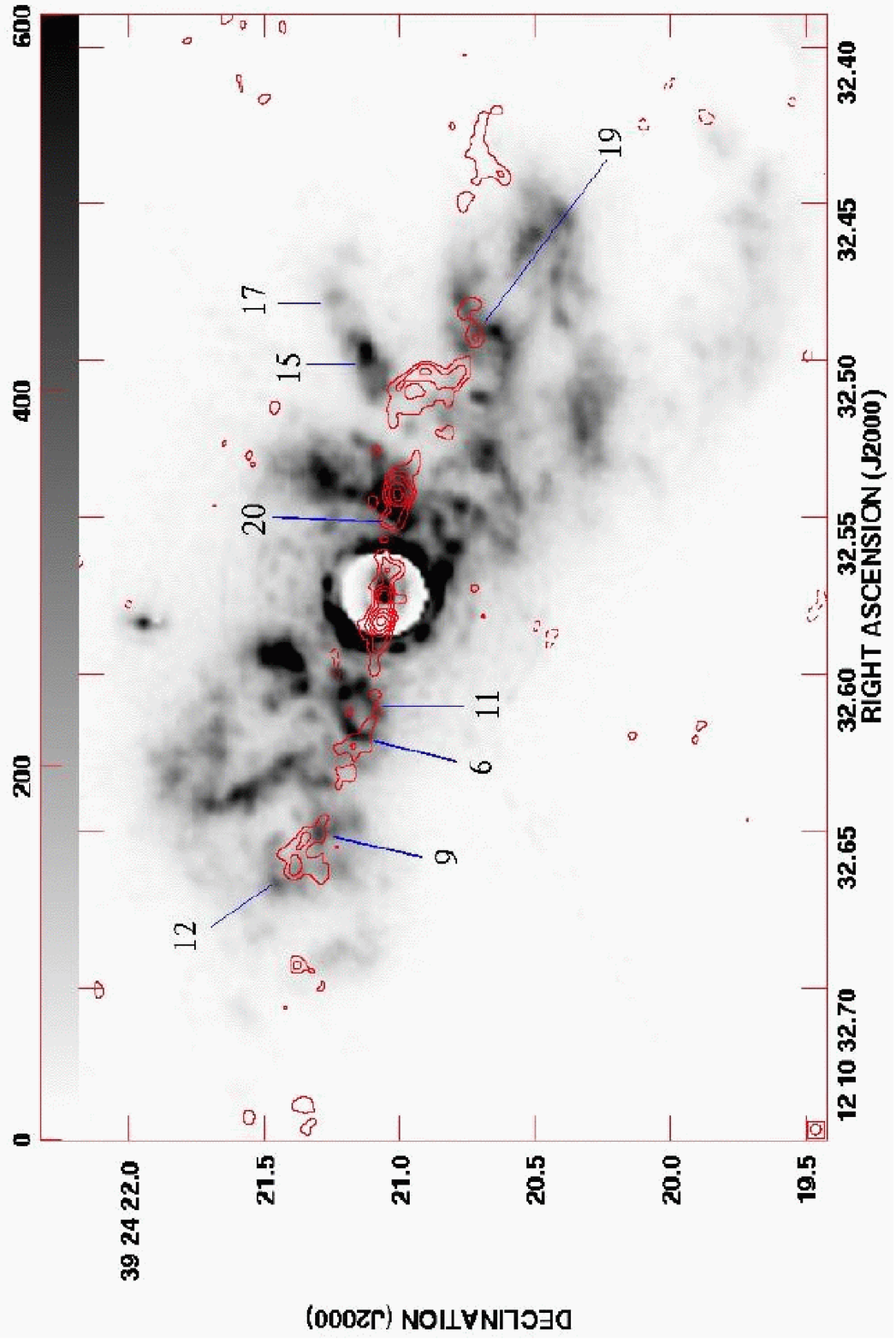}
\caption{FOC image in grey (Wing\'e et al. 1997) with contours of
radio jet (as in Figure \ref{lowresjet}) overlaid. [O{\sc iii}]-line
emitting clouds close to radio jet with high velocity dispersion are
labelled numerically following the convention of Hutchings et
al. (1998) and Kaiser et al. (2000).
\label{FOCjet}}
\end{figure*}

We are grateful to Jim Ulvestad, Andrew Wilson and John Porter for
useful discussions and Robin Catchpole for providing the FOC image
used in Figure \ref{FOCjet}.  We thank the referee, R. Antonucci, for
helpful suggestions that improved the paper. The National Radio
Astronomy Observatory is a facility of the National Science Foundation
operated under cooperative agreement by Associated Universities, Inc.
This research has made use of NASA's Astrophysics Data System Abstract
Service (ADS) and the NASA/IPAC Extragalactic Database (NED), which is
operated by the Jet Propulsion Laboratory, California Institute of
Technology, under contract with the National Aeronautics and Space
Administration. CGM acknowledges financial support from the Royal
Society.




\clearpage


\begin{table*}
\caption{Radio Component Designations (see also Figure \ref{lowresjet})}
\label{comps}
\begin{tabular}{lc} \\
\hline

Mundell et al. (1995)          & Ulvestad et al. 1998\\
		& and This Paper\\
\hline
C1 . . . . . . . . . . . . . . . . . . . . &A\\
C2 . . . . . . . . . . . . . . . . . . . . &B\\
C3 . . . . . . . . . . . . . . . . . . . . &C\\
C3-C4, eastern bridge. . . . . . . . .&X$^{\dagger}$\\
C4W. . . . . . . . . . . . . . . . . . .&D\\
C4E, peak. . . . . . . . . . . . . . . .&E\\
C4E, eastern `tail'. . . . . . . . . . . &F\\
C5, western extension. . . . . . . . .&G\\
C5 . . . . . . . . . . . . . . . . . . . . &H\\
C5, eastern extension . . . . . . . . . &I$^{\dagger}$\\
\hline
$^{\dagger}$New component identified in this paper
\end{tabular}\\
\end{table*}

\begin{table*}
\caption{Central velocity, width and mean column density for
absorption components detected at five locations across E+F, as shown
in Figure \ref{fivespec}. Systemic velocity is 998 \kms.}
\label{velcomps}
\begin{tabular}{cccc} \\
\hline
Box Label&	Component Velocity&	Width &			Mean \NH/T$_{\rm S}$\\
(see Fig. \ref{fivespec})& (Optical Heliocentric \kms)	&	(\kms)&		(10$^{19}$ \cmsq\ K$^{-1}$)\\
\hline
$\alpha$ &	 928.7	&		16	&		0.7\\
$\alpha$ &	 962.2	&		20	&		1.1\\
$\alpha$ &	1022.3	&		26	&		1.0\\
$\alpha$ &	1132.5	&		20	&		0.6\\
$\alpha$ &	1476.6	&		33	&		1.2\\
&&&\\
$\beta$ &		 825.2	&		20	&		1.8\\
$\beta$ &	 	925.4	&		20	&		1.8\\
$\beta$ &	 	978.9	&		20	&		2.1\\
$\beta$ &		1132.5 	&		20	&		0.9\\
$\beta$ &		1443.2	&		20	&		0.9\\
&&&\\
$\gamma$& 	 848.6	&		13	&		0.8\\
$\gamma$&	 982.2	&		20	&		2.0\\
$\gamma$&	1443.2	&		20	&		1.2\\
&&&\\
$\delta$&		1002.2	&		40	&		4.8\\
$\delta$&		1323.0	&		20	&		1.4\\
$\delta$&		1673.7	&		20	&		1.7\\
&&&\\
$\epsilon$&	 988.9	&		26	&		3.6\\
$\epsilon$&	1045.7	&		20	&		1.1\\
$\epsilon$&	1209.4	&		20	&		1.1\\
\hline
\end{tabular}\\
\end{table*}

\end{document}